\documentclass{PoS}
\usepackage{epsfig}
\title{HERA Collider Results}

\ShortTitle{HERA Collider Results}

\author{\speaker{A M Cooper-Sarkar}\\
        Oxford University\\
        E-mail: \email{amanda.cooper-sarkar@physics.ox.ac.uk}}

\abstract{The final results on the combination of all inclusive deep-inelastic
 scattering cross section data from the H1 and ZEUS experiments at HERA are 
reviewed. The parton distribution functions (PDFs) extracted from these data 
(HERAPDF2.0) and from HERA data on heavy flavour production and jet production 
are also presented (HERAPDF2.0Jets). The use of various heavy flavour schemes 
is compared and the sensitivity of the data to cuts on $Q^2$ is explored.
When jet data are included in the fits, a competitive value $\alpha_s(M_Z^2) = 0.1184 \pm 0.0016$ is extracted at NLO, excluding scale uncertainties}

\FullConference{XXIII International Workshop on Deep-Inelastic Scattering,\\
		27 April - May 1 2015\\
		Dallas, Texas}

\begin{document}

\section{Introduction}
Deep inelastic scattering (DIS) of electrons and positrons
on protons at HERA has been central to the exploration
of proton structure and quark-gluon interaction dynamics as
described by perturbative Quantum Chromo Dynamics (pQCD). 
HERA was operated at a centre-of-mass energy 
of up to $\sqrt{s} \simeq 320\,$GeV.
This enabled the two collaborations, H1 and ZEUS, to explore a large
phase space in Bjorken $x$, $x_{Bj}$, 
and negative four-momentum-transfer squared, $Q^2$.
Cross sections for neutral current (NC) interactions were
published for
$0.045 \leq Q^2 \leq 50000 $\,GeV$^2$
and  $6 \cdot 10^{-7} \leq x_{Bj} \leq 0.65$.

HERA was operated in two phases: HERA\,I, from 1992 to 2000, and HERA\,II, 
from 2002 to 2007. It was operated with an electron beam energy of
$E_e \simeq 27.5$\,GeV.
For most of HERA\,I and~II, the proton beam energy 
was $E_p = 920$\,GeV, resulting in the highest centre-of-mass energy of 
$\sqrt{s} \simeq 320\,$GeV.
During all of HERA running,
the H1 and ZEUS collaborations
collected total integrated luminosities of 
approximately 500\,pb$^{-1}$ each, 
divided about equally between $e^+p$ and $e^−p$ scattering. 
The data presented here is the final combination of HERA inclusive data
based on all published H1 and ZEUS measurements corrected to zero beam polarisation. This includes data taken 
with proton beam energies of
$E_p = 920$, 820,  575 and 460\,GeV
corresponding to 
at $\sqrt{s}\simeq$\,320, 300, 251 and 225\,GeV.

\begin{figure}[tbp]
\vspace{-0.5cm} 
%\vspace*{5pt}
\centerline{
\epsfig{figure=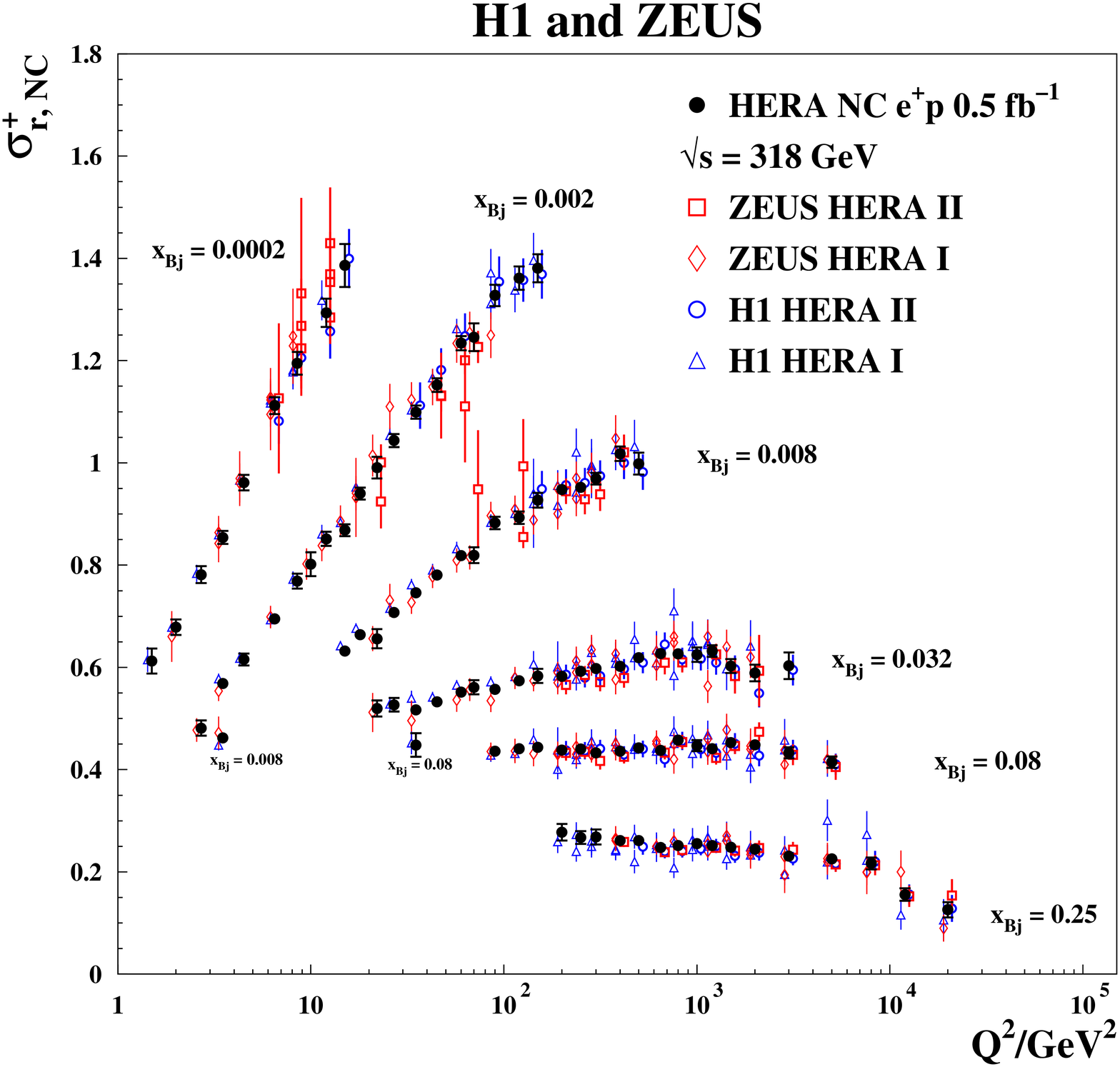,width=0.7\linewidth}}
\centerline{
\epsfig{figure=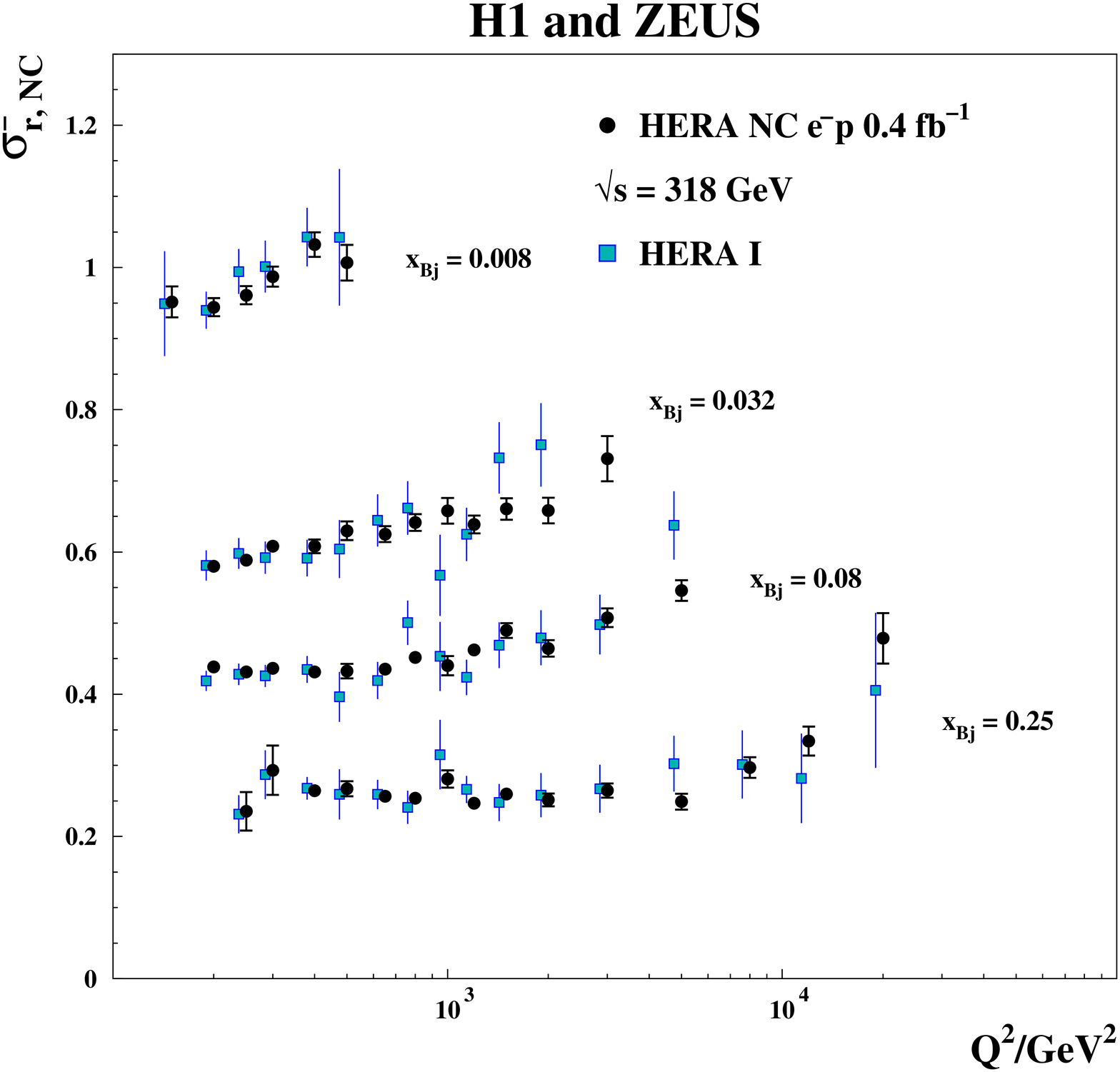,width=0.7\linewidth}}
\caption {HERA combined NC $e^+p$ reduced 
cross sections as a function of 
$Q^2$ for selected $x_{\rm Bj}$-bins compared to the individual 
H1 and ZEUS data(top); and HERA combined NC $e^-p$ reduced cross sections compared to the to the HERA-I combination (bottom).
}
\label{fig:d15039f4f7}
\end{figure}

\begin{figure}[tbp]
\vspace{-0.5cm} 
%\vspace*{5pt}
\centerline{
\epsfig{figure=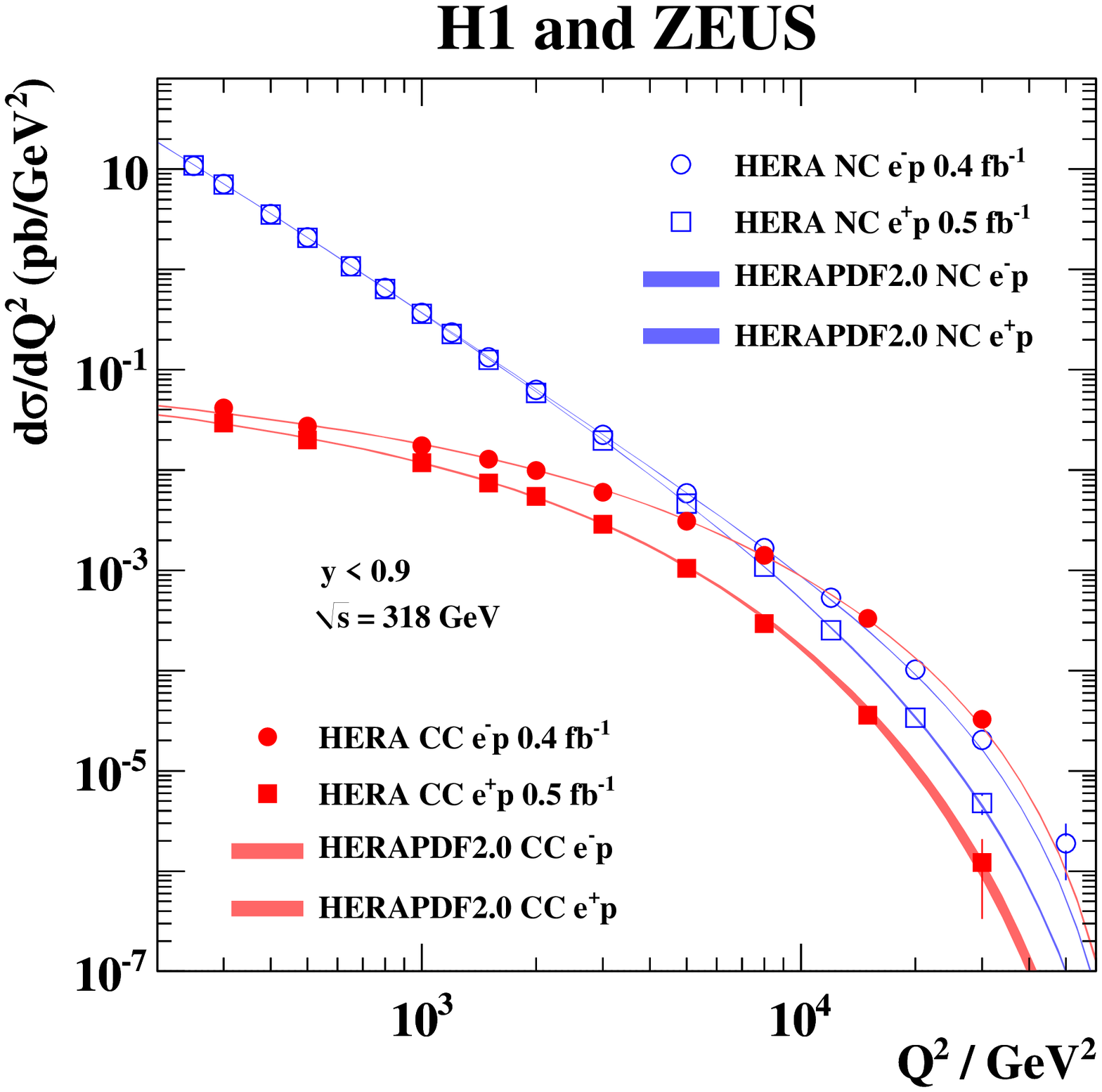,width=0.7\linewidth}}
\centerline{
\epsfig{figure=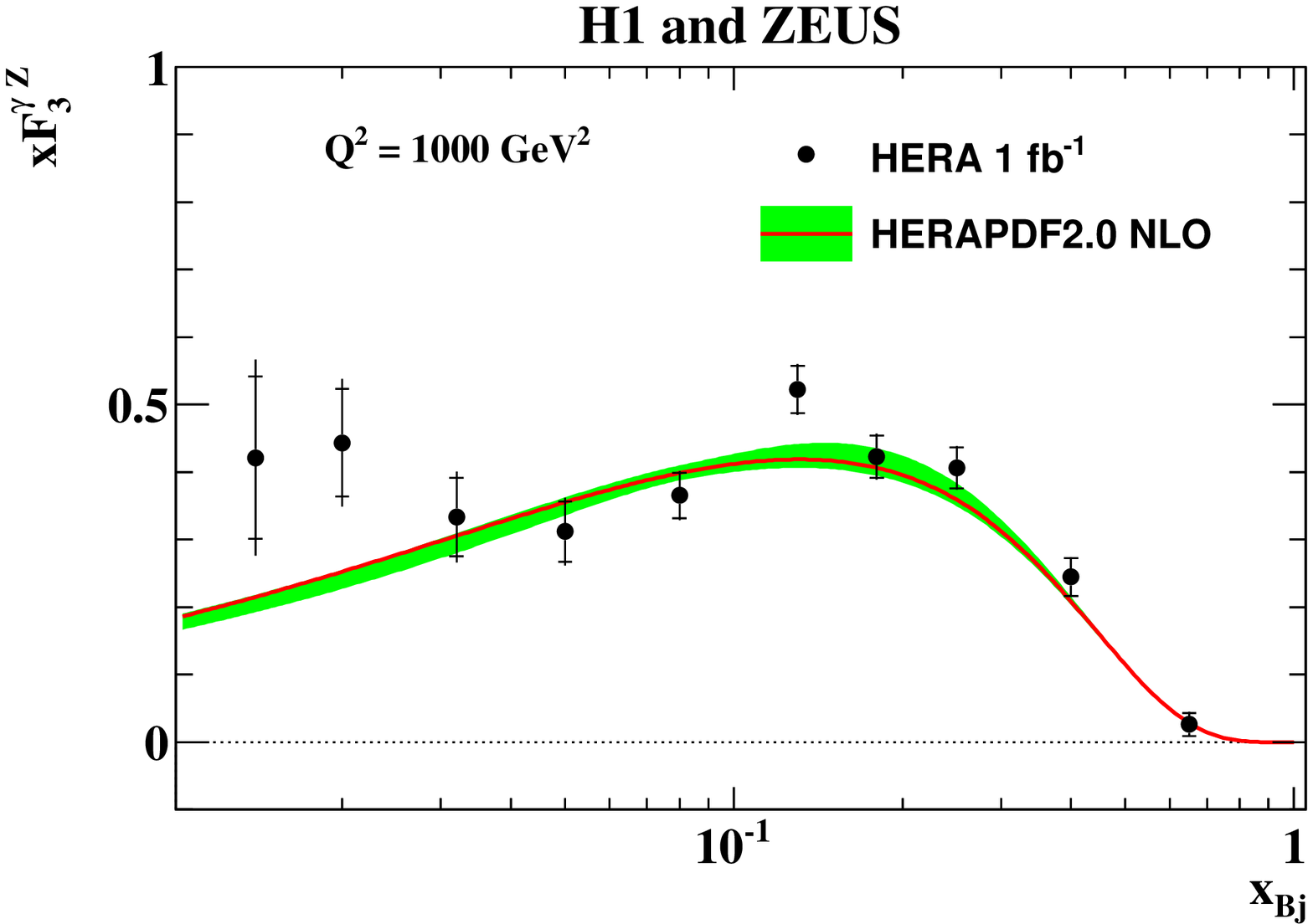,width=0.7\linewidth}}
\caption {NC and CC  $e^-p$ and $e^+p$ cross sections (top)
The structure function $xF_3^{\gamma Z}$ at $Q^2=1000\,$GeV$^{2}$ (bottom). 
The data are compared with the prediction from HERAPDF2.0 NLO.
}
\label{fig:d15039f74f78}
\end{figure}

The combination was performed using the 
package HERAverager~\cite{HERAverager} and the pQCD analysis using
HERAFitter~\cite{HERAFitter}. 
The correlated systematic uncertainties
and global normalisations were treated 
such that one coherent data set was obtained. 
The combination leads to a significantly reduced uncertainty compared to the 
orginal inputs and compared to the previous combination of HERA-I data, 
particularly in the electron sector, see Fig.~\ref{fig:d15039f4f7}.
The combined data demonstrate electroweak unification beautifully and allow an extraction of $xF_3^{\gamma Z}$, see Fig.~\ref{fig:d15039f74f78}

Within the framework of pQCD, 
the proton is described
by parton distribution functions (PDFs) which provide probabilities
for a particle
to scatter off partons, gluons or quarks, 
carrying the fraction $x$ of the proton momentum.
Perturbative QCD determines the evolution of the PDFs to any scale once 
they are provided at a starting scale. The name HERAPDF stands for a pQCD 
analysis, within the
DGLAP formalism, to determine the PDFs at the starting scale by fitting the $x_{Bj}$ and $Q^2$ dependences of the combined HERA NC and CC DIS
cross sections. The name
HERAPDF2.0 refers to this analysis based on
the newly combined inclusive DIS cross sections from all of HERA~I and HERA~II.
The strength of the HERAPDF approach is that 
one coherent high-precision data set containing NC and CC cross sections
is used as input.
The newly combined data entering the HERAPDF2.0 analysis span
four orders of magnitude in $Q^2$ and $x_{Bj}$.
The availability of precision NC and CC cross sections 
over such a large large phase space allows HERAPDF to be based on
$ep$ scattering data only and makes HERAPDF independent 
of any nuclear corrections. 
The difference between the NC $e^+p$ and $e^-p$ cross sections 
at high $Q^2$, together with the high-$Q^2$ CC data,
constrain the valence quark distributions.
The CC data also constrain
the down sea-quark distribution in the proton without assuming
isospin symmetry.     
The lower-$Q^2$ NC data 
constrain the low-$x$ sea-quark distributions.
The precisely measured $Q^2$ variations 
of the DIS cross sections
in different bins of $x_{Bj}$
constrain the gluon distribution. Measurement of cross sections at 
different beam energies constrains the longitudinal structure function, $F_L$, 
and thus provides independent information on the gluon distribution.
 
The consistency of the input data allowed the determination of the 
experimental uncertainties on the HERAPDF2.0 parton distributions  
using rigorous statistical methods.  
Uncertainties resulting from model assumptions
and from the choice of the parameterisation of the PDFs 
are considered separately. 

Both H1 and ZEUS also published charm-production cross sections,
which were combined and analysed previously, as well
as jet-production cross sections.
These data were included to obtain a variant HERAPDF2.0Jets.
The inclusion of jet cross-sections made it possible to simultaneously
determine the PDFs and the strong coupling constant $\alpha_s(M_Z^2)$.
Full details of the analysis are given in ref.~\cite{thepaper}.

\section{HERAPDF2.0 and its variations}
\begin{figure}[tbp]
\vspace{-0.5cm} 
%\vspace*{5pt}
\centerline{
\epsfig{figure=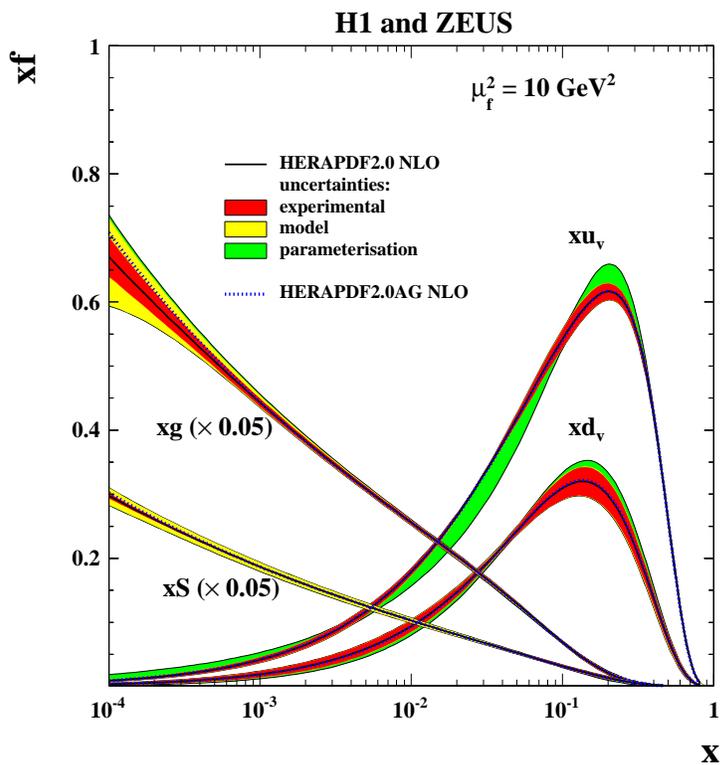,width=0.7\linewidth}}
\centerline{
\epsfig{figure=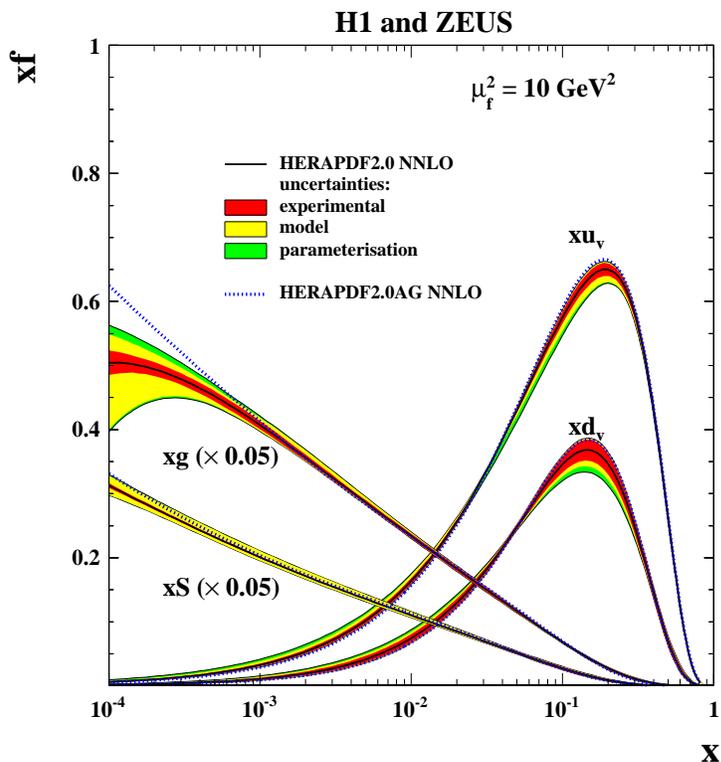,width=0.7\linewidth}}
\caption {The parton distribution functions of 
HERAPDF2.0 NLO(top) and NNLO(bottom), $xu_v$, $xd_v$, $xS=2x(\bar{U}+\bar{D})$, $xg$, 
at $\mu_{\rm f}^{2} = 10\,$GeV$^{2}$.
The gluon and sea distributions are scaled down 
by a factor of $20$.
The experimental, model and parameterisation 
uncertainties are shown 
separately. 
The dotted lines represent HERAPDF2.0AG NLO with an alternative
gluon parameterisation.
}
\label{fig:d15039f21f23}
\end{figure}

\begin{figure}[tbp]
\vspace{-0.5cm} 
%\vspace*{5pt}
\centerline{
\epsfig{figure=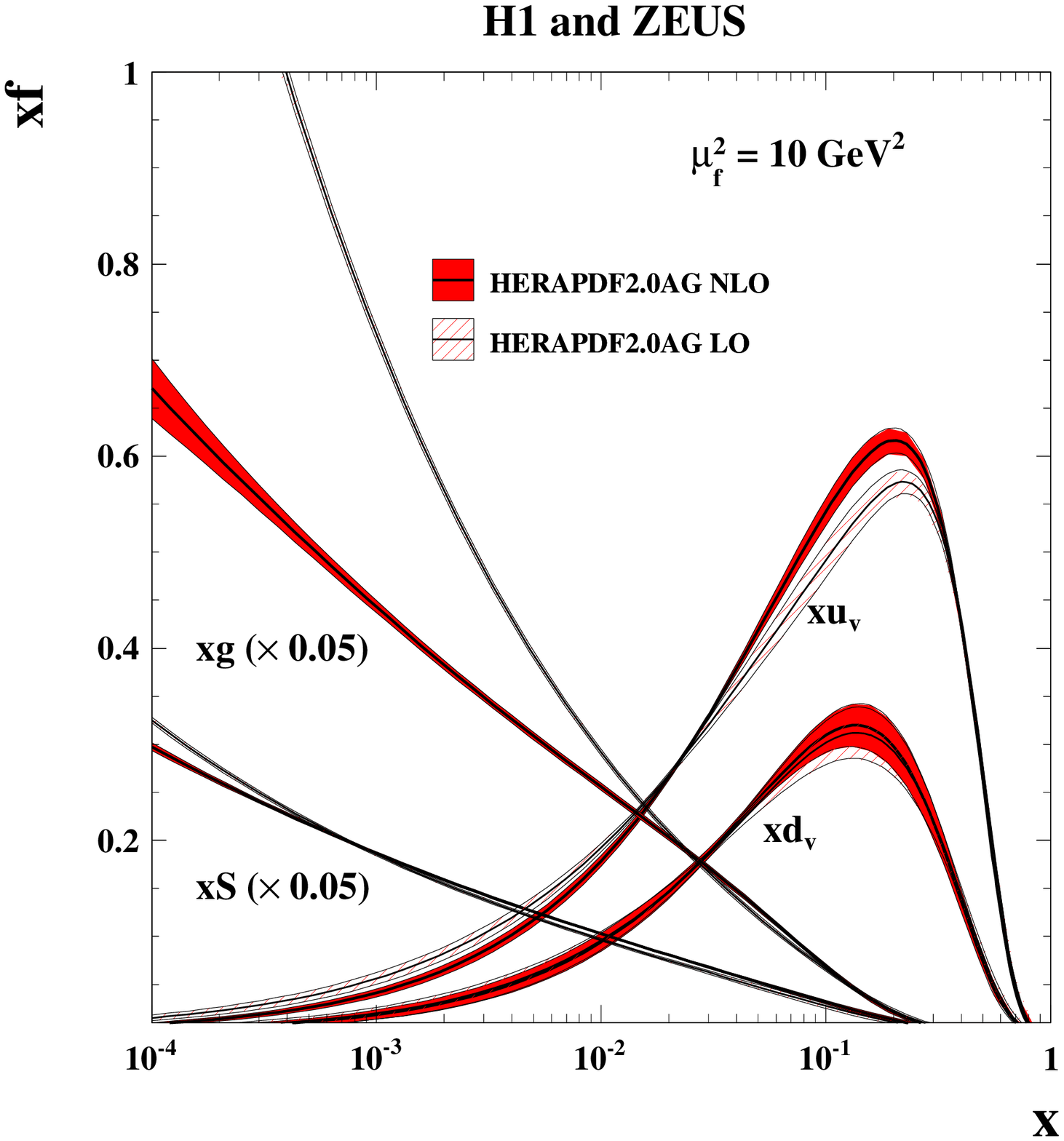,width=0.7\linewidth}}
\caption {HERAPDF2.0 PDFs at LO and NLO are compared using experimental uncertaities only. The alternative gluon form of the parametrisation is used for both.
}
\label{fig:d15039f26}
\end{figure}

Fig.~\ref{fig:d15039f21f23} shows summary plots at $\mu_f=10$GeV$^2$ of the 
valence, total Sea and 
gluon PDFs for HERAPDF2.0 analysed at NLO and at NNLO, for the standard cut 
$Q^2 >3.5$GeV$^2$. The experimental uncertainties are shown in red. Model 
uncertainties are shown in yellow. These are due to variation of the central 
choices for: the $Q^2$ cut; the values of the pole-masses of the charm and beauty 
quarks; the fractional contribution and the shape of the strange-PDF. HERA data on charm and beauty 
production determine the central choices of the heavy quark masses and their model
variations. Parametrization uncertainties are shown in green. These are due 
to: variation of the starting scale; addition of extra parameters. The central 
choice of parametrization is determined as usual for HERAPDF analyses by saturation of the 
$\chi^2$, but the addition of extra parameters sometimes results in close-by 
but distinct minima. Additionally seen on these figures is the result of an 
alternative gluon parametrisation HERAPDF2.0AG, for which the gluon must be 
positive definite for all $Q^2$ above the starting scale. These PDFs are similar
 in $\chi^2$ to the standard ones at NLO but are disfavoured at NNLO. 
An LO set is also available using the alternative gluon parametrisation. 
It is shown compared to the NLO set in Fig.~\ref{fig:d15039f26}. The standard fits use a 
value of the strong coupling constant, $\alpha_s(M_Z^2)=0.118$, at NLO and NNLO, and, 
$\alpha_s(M_Z^2)=0.130$, at LO, but sets using a range of values from 
$0.110$ to $0.130$ are also available.

\begin{figure}[tbp]
\vspace{-0.5cm} 
%\vspace*{5pt}
\centerline{
\epsfig{figure=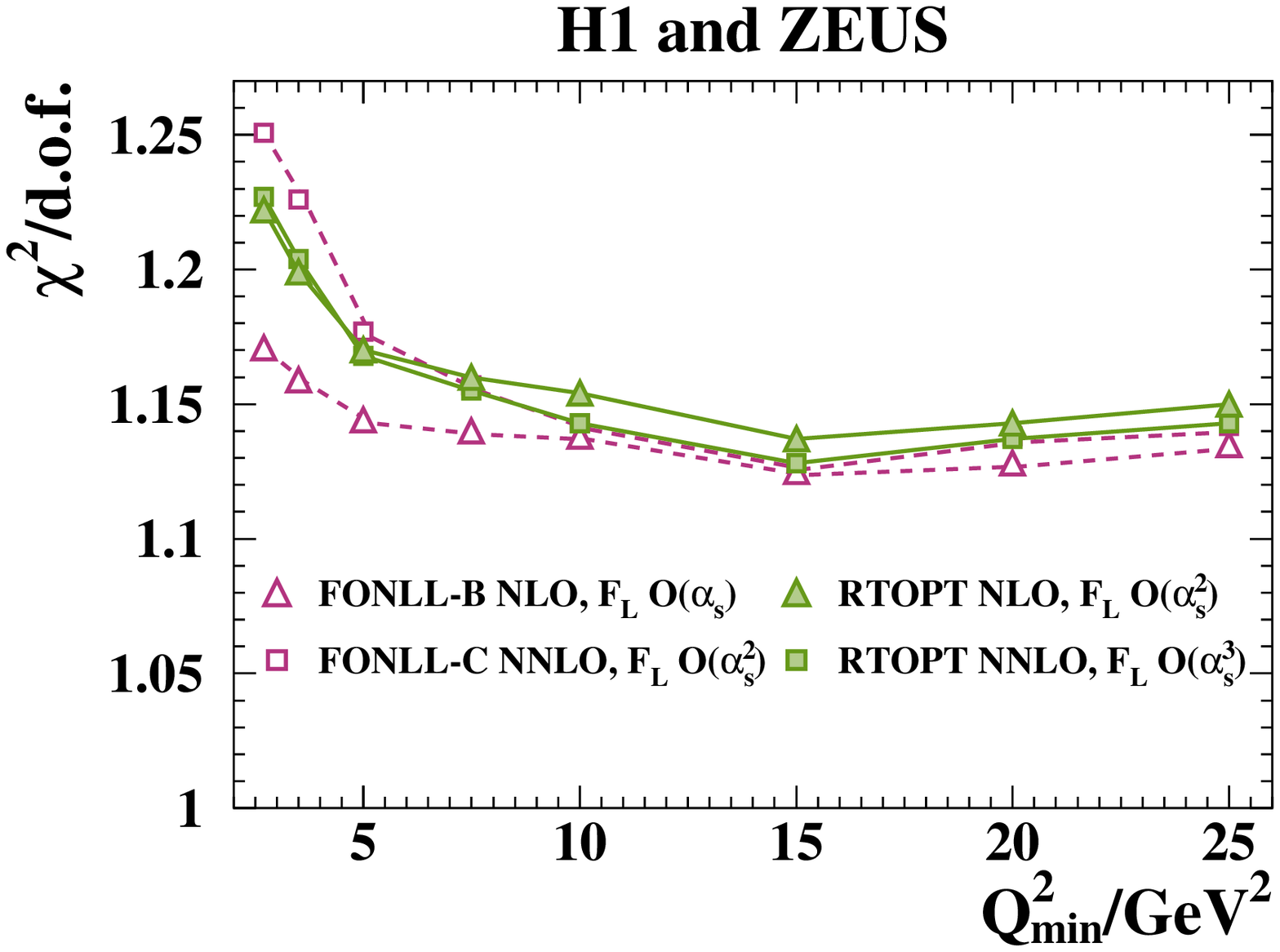,width=0.7\linewidth}}
\centerline{
\epsfig{figure=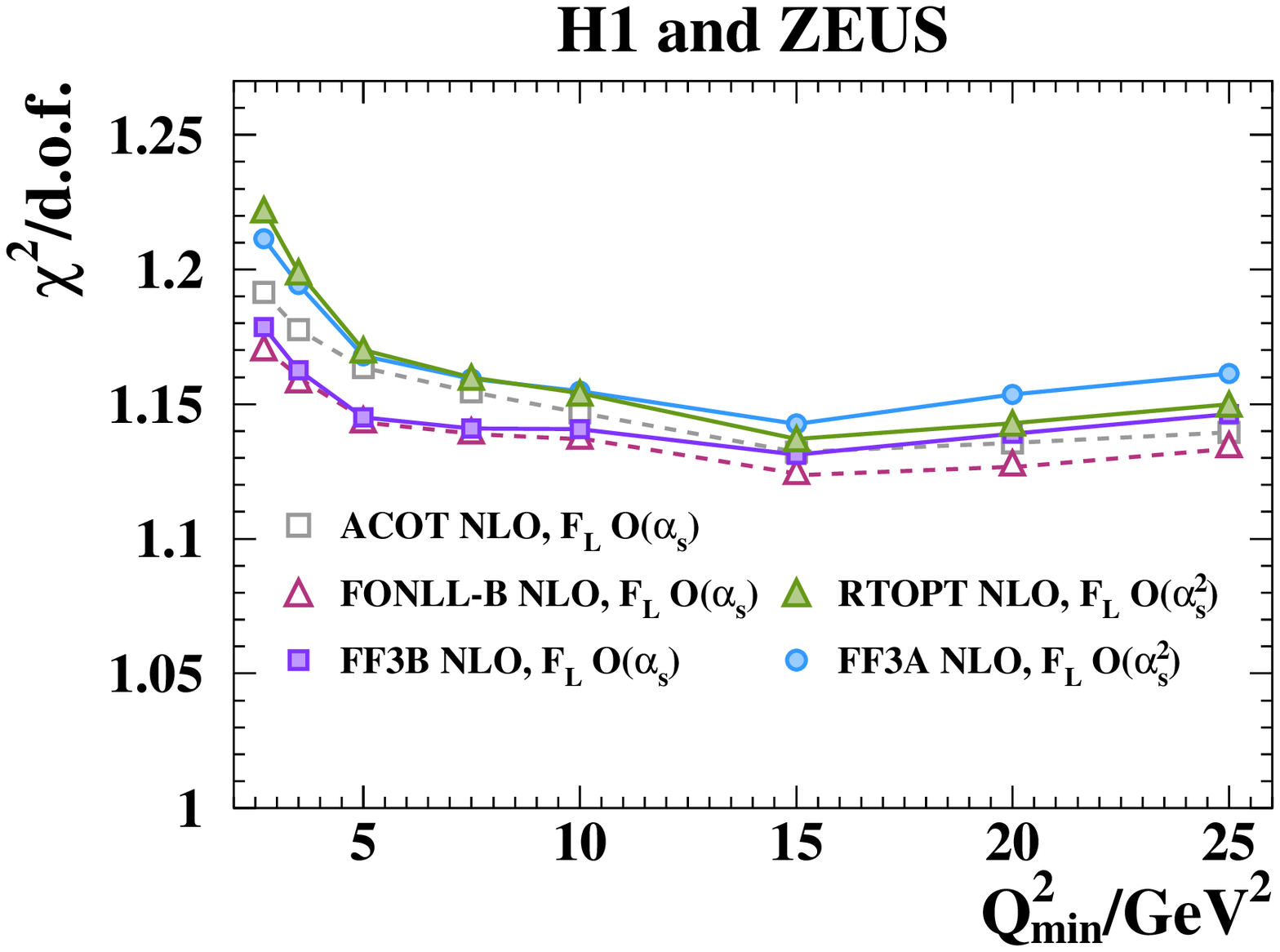,width=0.7\linewidth}}
\caption {The dependence of $\chi^2/d.o.f$ on $Q^2_{min}$ for HERAPDF2.0 fits: 
(top) using the RTOPT and FONLL schmes at NLO and NNLO; (bottom) using RTOPT, 
ACOT and FONLL-B schemes and fixed flavour number schemes at NLO.} 
\label{fig:d15039f20af20b}
\end{figure}

A more extreme variation of the $Q^2$ cut, $Q^2 > 10$GeV$^2$, has also been 
considered, resulting in the HERAPDF2.0HiQ2 PDFs, since it was observed that 
the $\chi^2$ per degree of freedom of 
the fit decreases steadily until $Q^2 > 10$GeV$^2$. This is true for both NLO and 
NNLO fits and it is
true independent of the heavy quark scheme used to analyse the data, see Fig.~\ref{fig:d15039f20af20b}. In fact   
it depends mostly on the order to which $F_L$ is evaluated. The fits do not 
favour the evaluation of $F_L$ to $O(\alpha_s^2)$. It is also somewhat counter-intuitive that the $\chi^2$ is not improved when going from NLO to NNLO within the same scheme, 
the fits do not favour the faster 
NNLO evolution.

\begin{figure}[tbp]
\vspace{-0.5cm} 
%\vspace*{5pt}
\centerline{
\epsfig{figure=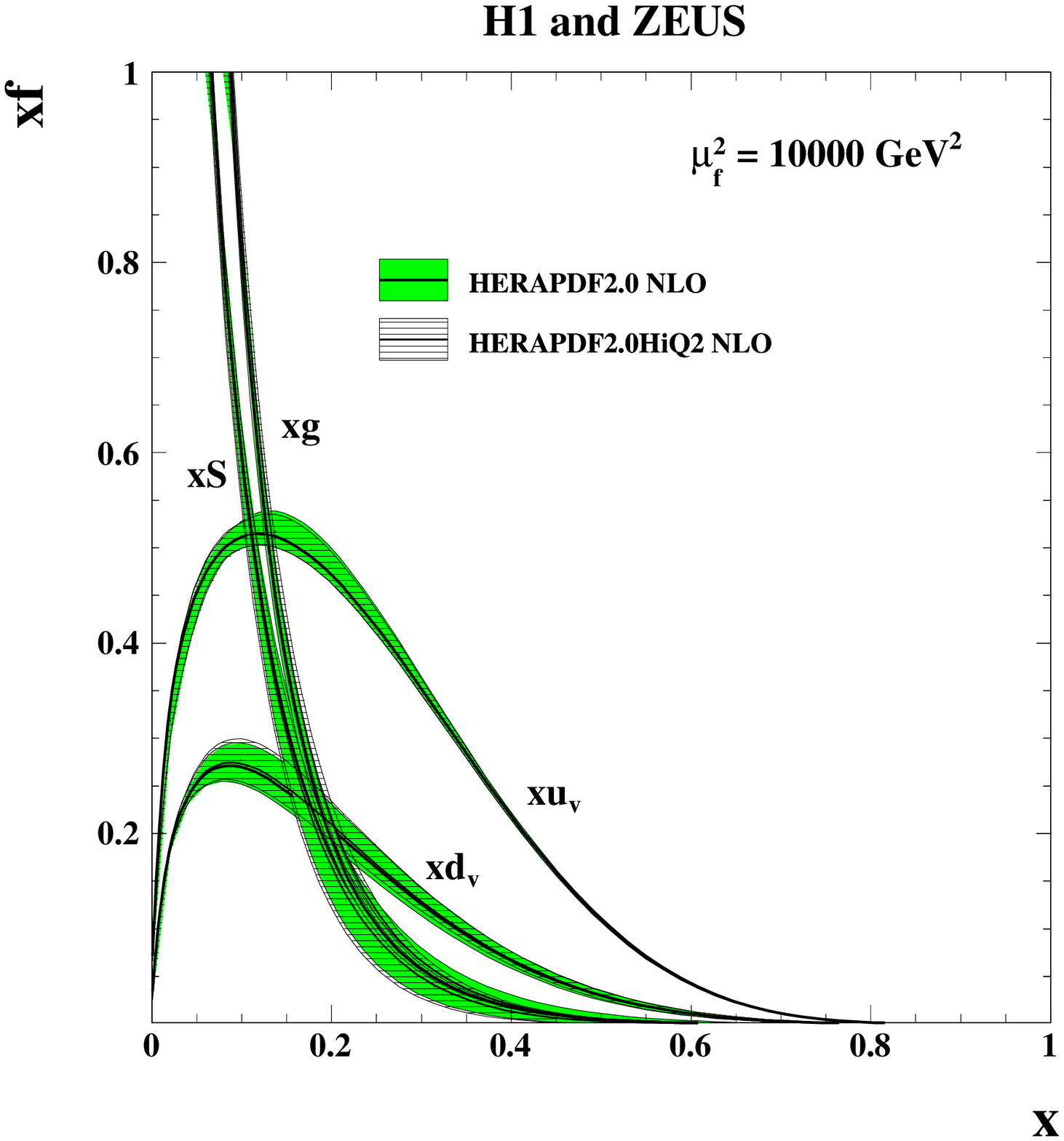,width=0.7\linewidth}}
\centerline{
\epsfig{figure=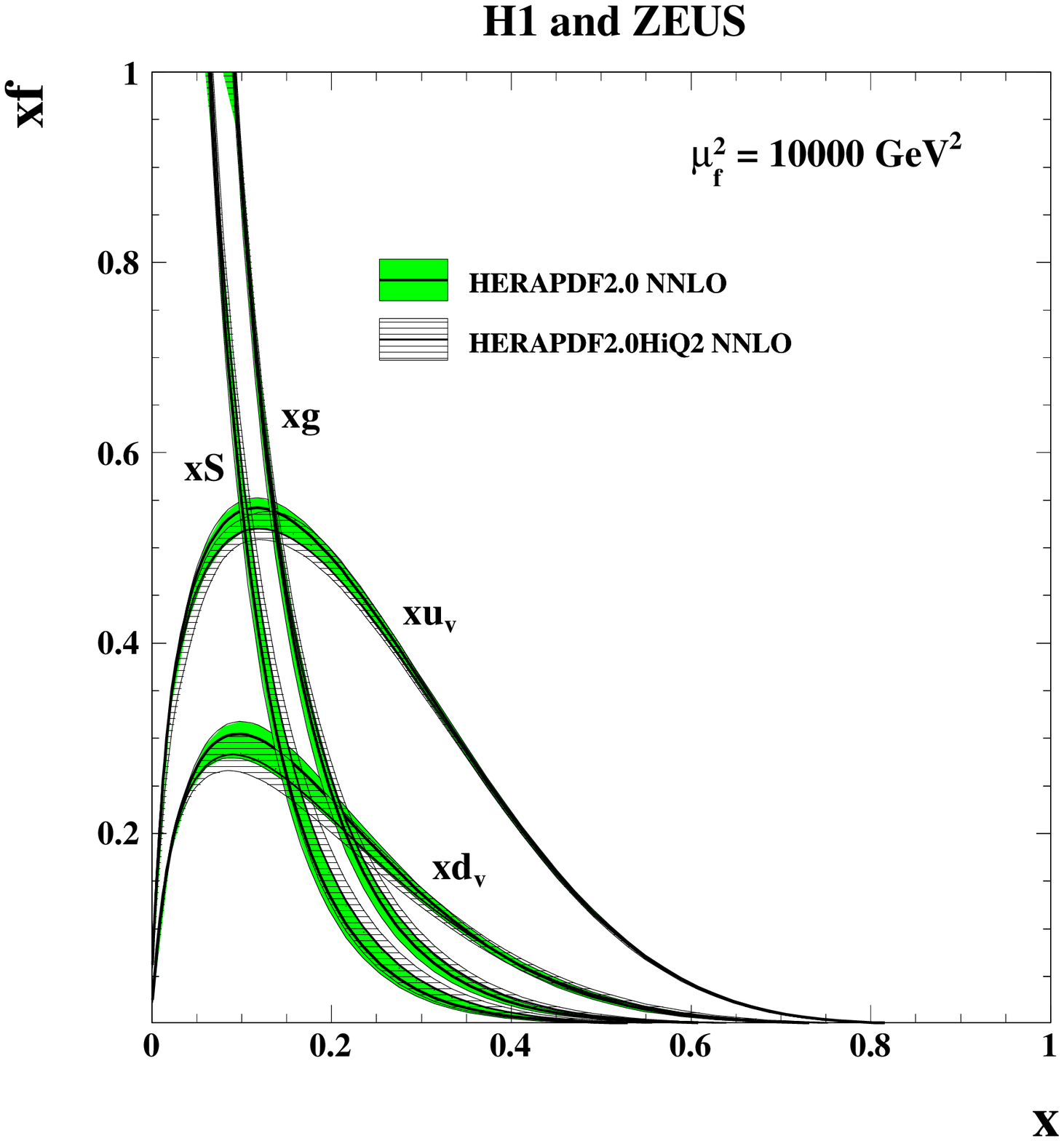,width=0.7\linewidth}}
\caption {The parton distribution functions of 
HERAPDF2.0, $xu_v$, $xd_v$, $xS=2x(\bar{U}+\bar{D})$, $xg$, 
at $\mu_{\rm f}^{2} = 10000\,$GeV$^{2}$ with $Q^{2}_{\rm min} = 3.5$\,GeV$^{2}$
compared to the PDFs of HERAPDFHiQ2 with $Q^2_{\rm min}=10$\,GeV$^2$.
Top for the NLO fits; bottom for the NNLO fits
}
\label{fig:d15039f55bf57b}
\end{figure}

 HERA kinematics are such that low $Q^2$ is also low $x$. Thus HERAPDF2.0HiQ2 PDFs are 
used to assess any bias resulting from the inclusion of low-$Q^2$, low-$x$ data 
which might require analysis beyond the DGLAP formalism, such as: resummation
of $ln(1/x)$ terms, non-linear evolution equations and non-perturbative 
effects. Figs.~\ref{fig:d15039f55bf57b} show that there is no bias at high scale 
due to the inclusion of the lower $Q^2$ data. 

\begin{figure}[tbp]
\vspace{-0.5cm} 
%\vspace*{5pt}
\centerline{
\epsfig{figure=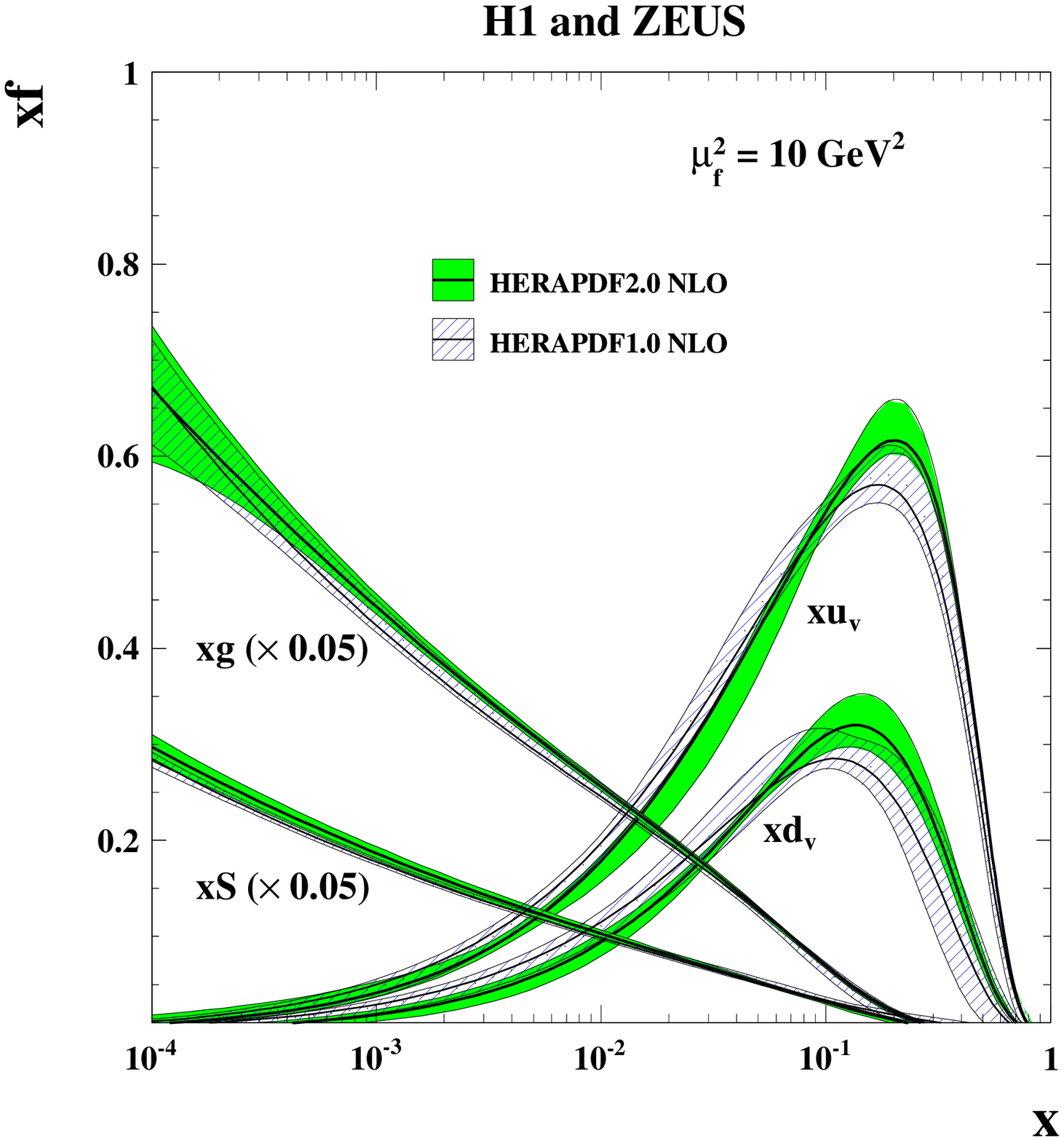,width=0.7\linewidth}}
\centerline{
\epsfig{figure=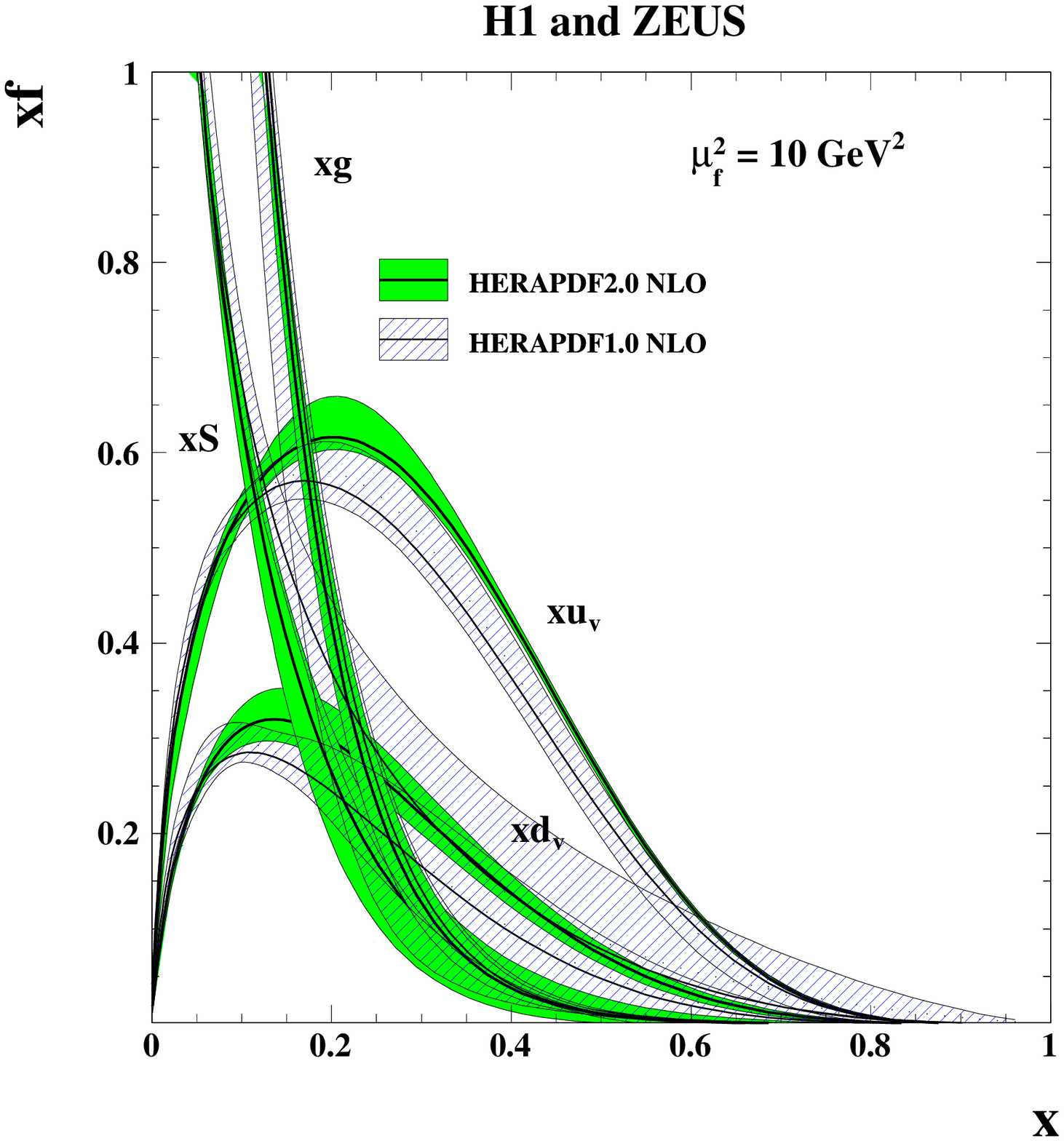,width=0.7\linewidth}}
\caption {The parton distribution functions of
HERAPDF2.0 NLO, 
$xu_v$, $xd_v$, $xS=2x(\bar{U}+\bar{D})$, $xg$,
at $\mu_{\rm f}^{2}=10\,$GeV$^{2}$ 
compared to HERAPDF1.0NLO on log (top)
and linear (bottom) scales. 
}
\label{fig:d15039f47af47b}
\end{figure}

\begin{figure}[tbp]
\vspace{-0.5cm} 
%\vspace*{5pt}
\centerline{
\epsfig{figure=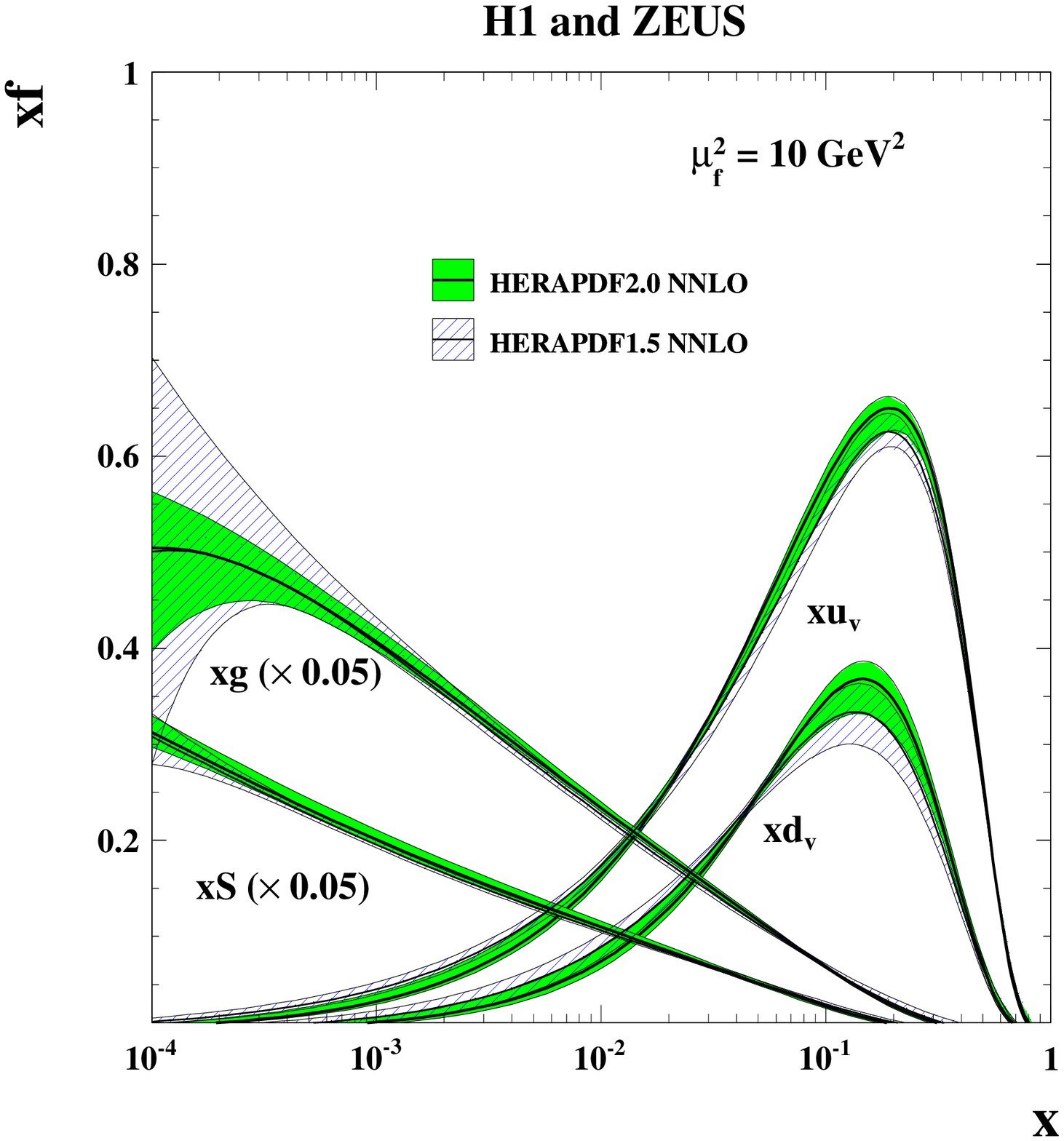,width=0.7\linewidth}}
\centerline{
\epsfig{figure=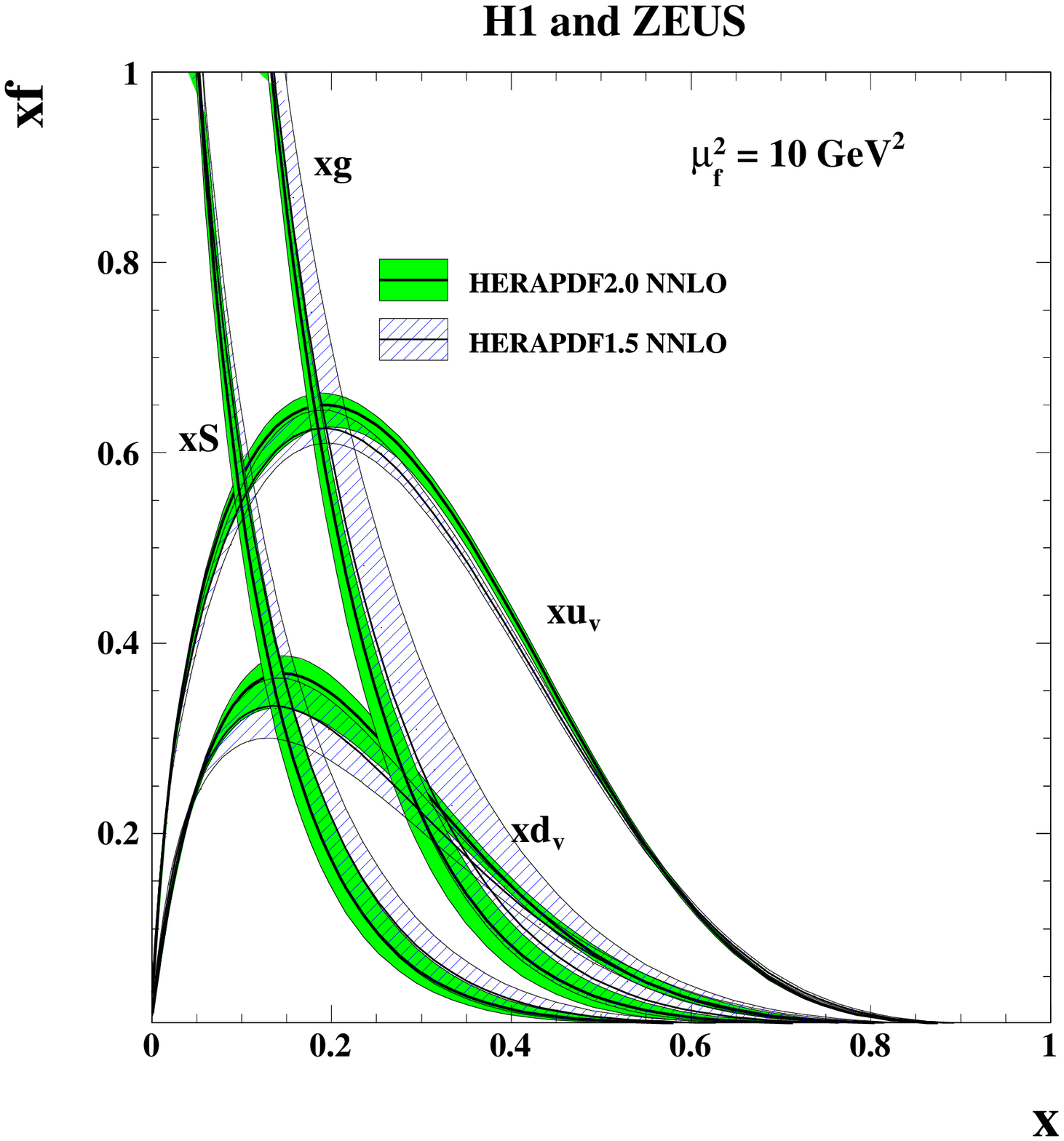,width=0.7\linewidth}}
\caption {The parton distribution functions of
HERAPDF2.0 NNLO, 
$xu_v$, $xd_v$, $xS=2x(\bar{U}+\bar{D})$, $xg$,
at $\mu_{\rm f}^{2}=10\,$GeV$^{2}$ 
compared to HERAPDF1.5 NNLO on log (top)
and linear (bottom) scales. 
}
\label{fig:d15039f49af49b}
\end{figure}

Figs~\ref{fig:d15039f47af47b} compare the HERAPDF2.0NLO fit to the
 HERAPDF1.0NLO fit. One can see the reduction in the high-x uncertainties and 
the fact that the high-$x$ Sea is now much less hard. 
Figs.~\ref{fig:d15039f49af49b} make the same comparison for the HERAPDF2.0NNLO and the HERAPDF1.5NNLO fit. 
Again the high-x uncertainty is reduced and in particular, the high-x gluon has a much 
reduced uncertainty band and its central value moves towards the lower end of the 
HERAPDF1.5 uncertainty band.

\begin{figure}[tbp]
\vspace{-0.5cm} 
%\vspace*{5pt}
\centerline{
\epsfig{figure=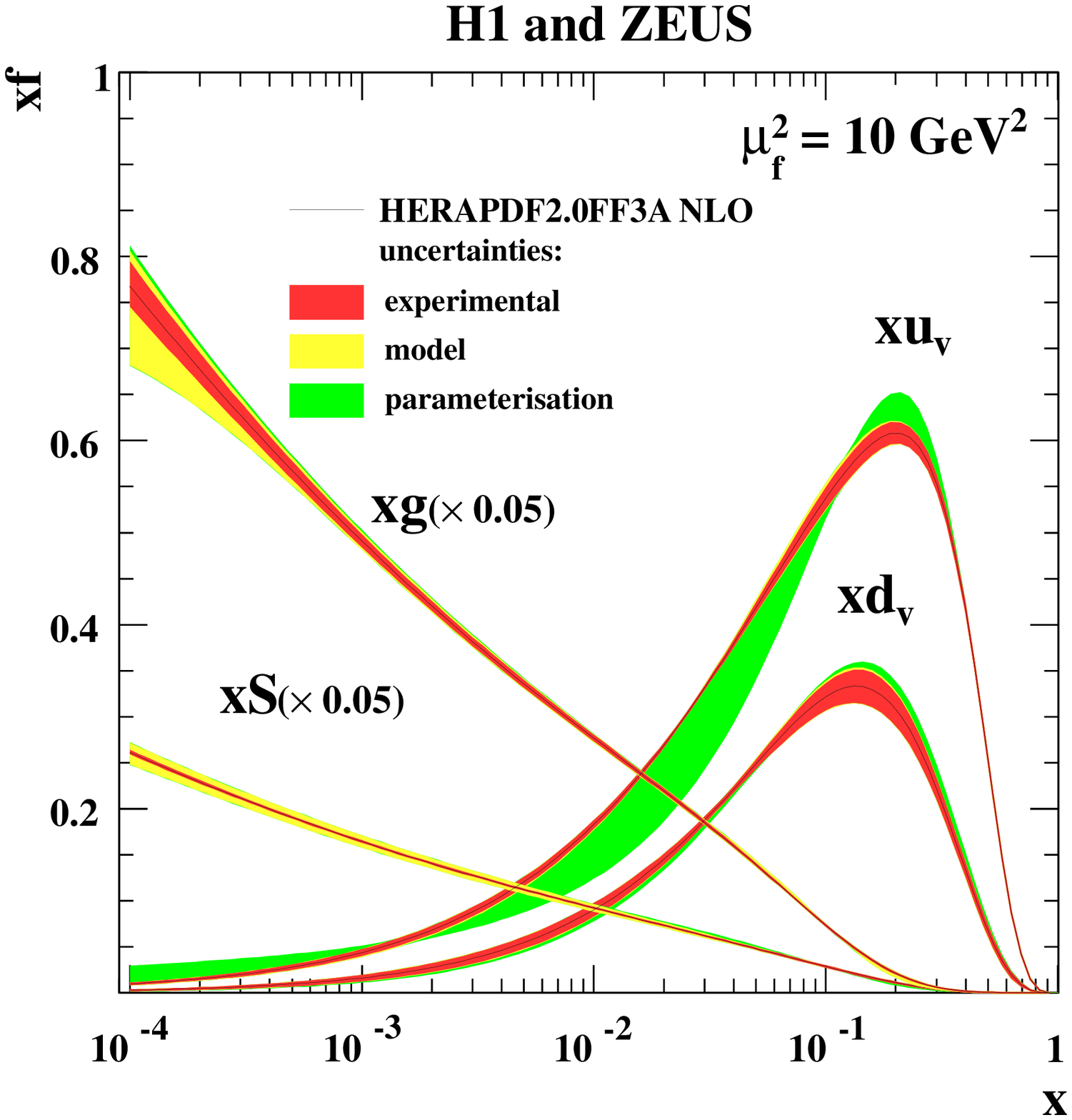,width=0.7\linewidth}}
\centerline{
\epsfig{figure=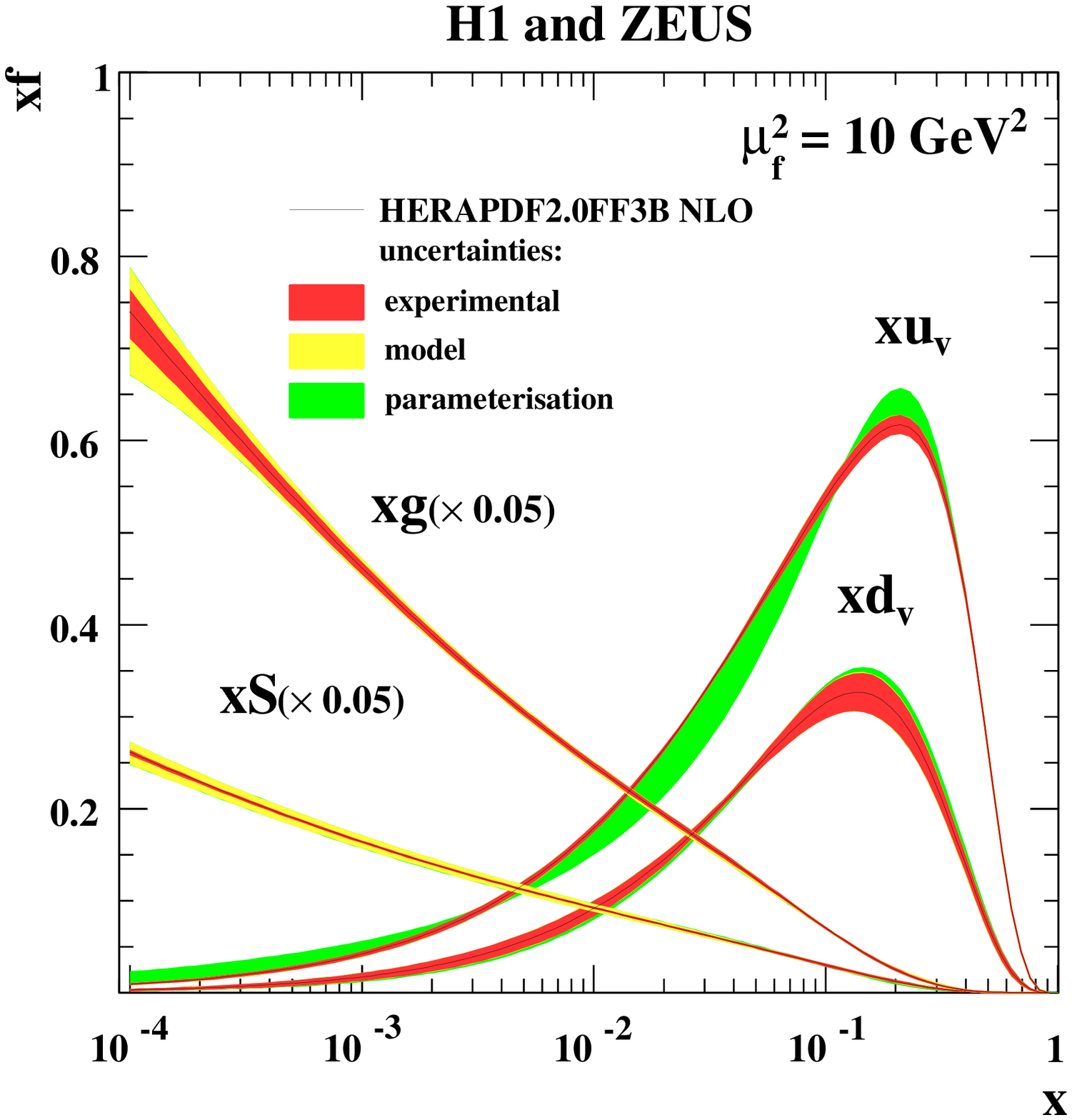,width=0.7\linewidth}}
\caption {The parton distribution functions of HERAPDF2.0FF3A NLO
and HERAPDF2.0FF3B NLO,
$\textit{xu}_{\textit{v}}$,$\textit{xd}_{\textit{v}}$,$\textit{xS=2x(}\bar{U
}+\bar{D}$),$ \textit{xg}$, 
at $\mu_{\rm f}^{2}$ = 10\,GeV$^{2}$.
% with $Q^{2}_{\rm min} = 3.5$\,GeV$^{2}$.
The experimental, model
and parameterisation uncertainties are shown separately.
}
\label{fig:d15039f60af60b}
\end{figure}

Two sets of PDFs using a fixed flavour number scheme have been extracted, as 
shown in Fig~\ref{fig:d15039f60af60b}. These differ from each other in three respects: the order at 
which $F_L$ is evaluated $O(\alpha_s^2)$ (FF3A), $O(\alpha_s)$ (FF3B); whether or not $\alpha_s$ runs with 3-flavours (FF3A)
 or with variable flavour (FF3B); and the use of pole masses (FF3A) or current masses (FF3B).

\begin{figure}[tbp]
\vspace{-0.5cm} 
%\vspace*{5pt}
\centerline{
\epsfig{figure=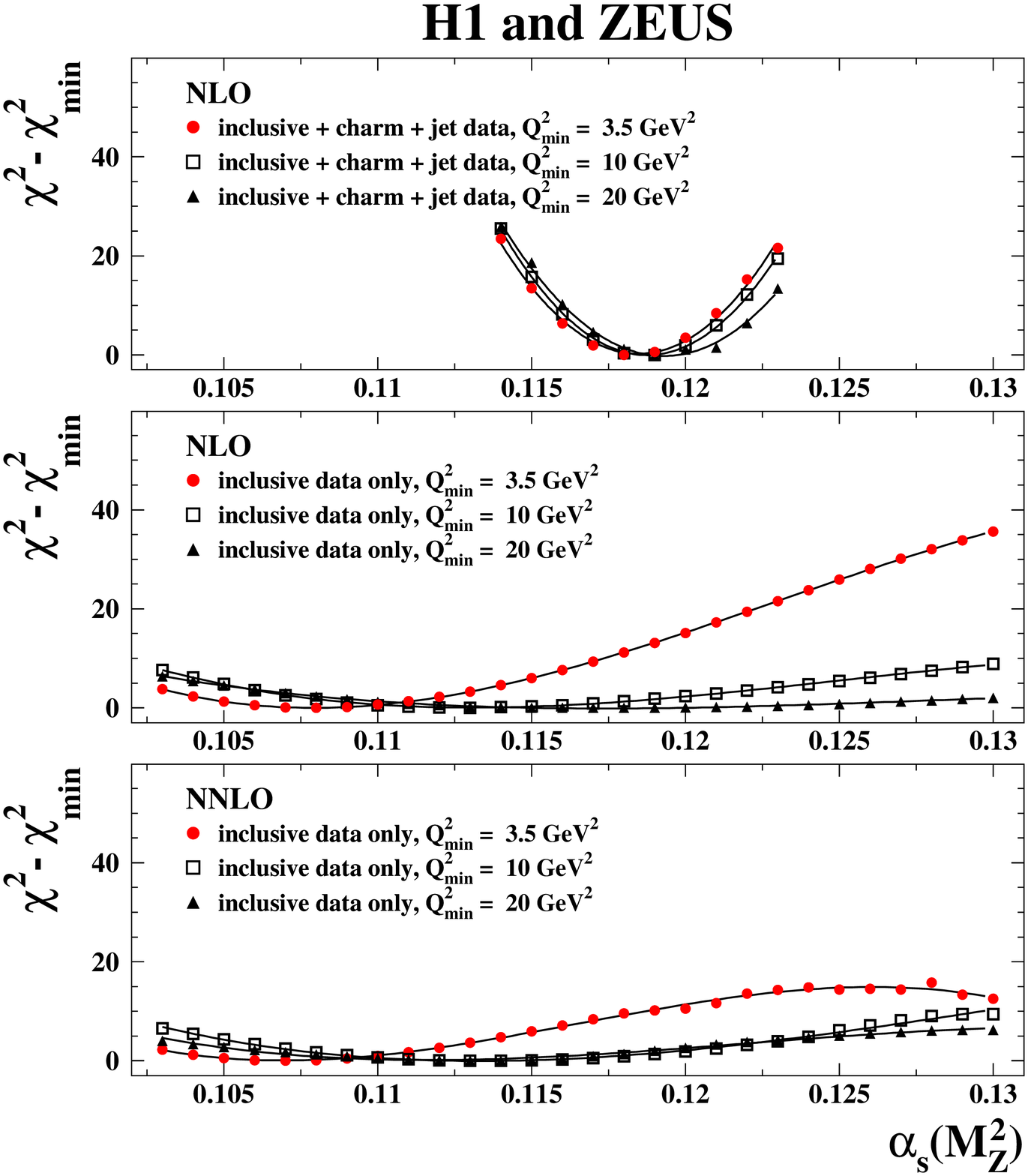,width=0.7\linewidth}}

\caption {$\Delta \chi^2$ vs. $\alpha_s(M_Z^2)$ for pQCD fits with different $Q^2_{\rm min}$ using
 data on (a) inclusive, charm and jet production at NLO,
(b) inclusive $ep$ scattering only at NLO, and
(c) inclusive $ep$ scattering only at NNLO.
}
\label{fig:d15039f65}
\end{figure}

\begin{figure}[tbp]
\vspace{-0.5cm} 
%\vspace*{5pt}
\centerline{
\epsfig{figure=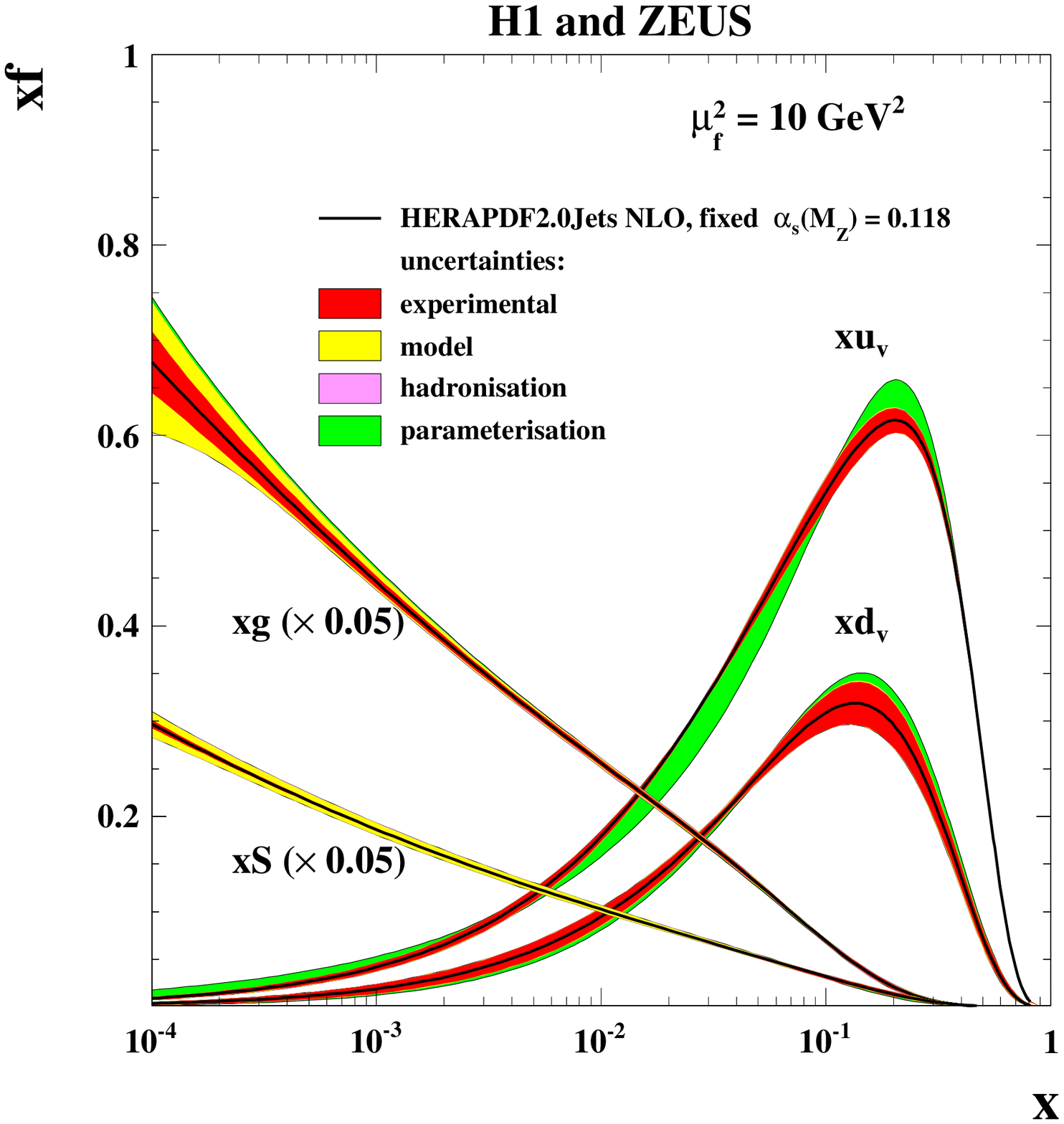,width=0.7\linewidth}}
\centerline{
\epsfig{figure=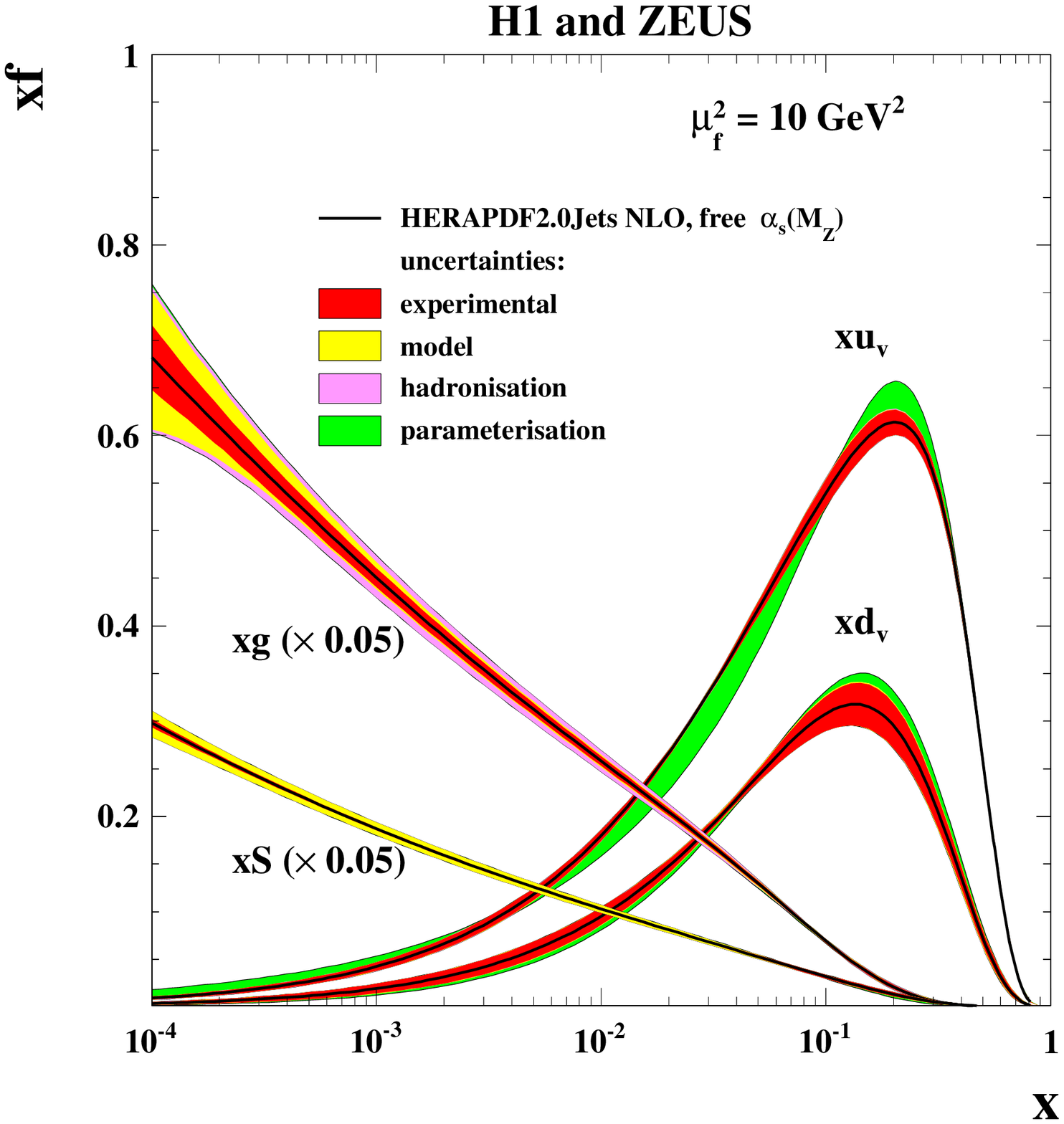,width=0.7\linewidth}}
\caption {The parton distribution functions of
HERAPDF2.0Jets NLO, $xu_v$, $xd_v$, $xS=2x(\bar{U}+\bar{D})$, $xg$, 
at $\mu_{\rm f}^{2} = 10\,$GeV$^{2}$ 
%with $Q^{2}_{\rm min} = 3.5$\,GeV$^{2}$
with fixed $\alpha_s(M_Z^2)=0.118$ (top) and free $\alpha_s(M_Z^2)$ (bottom).
The experimental, model and parameterisation 
uncertainties are shown separately. 
The hadronisation uncertainty is also included, but it is
only visible for the fit with free $\alpha_s(M_Z^2)$.
}
\label{fig:d15039f66af66b}
\end{figure}

Heavy flavour data from the charm combination has also been added to the fit, 
but it does not make much difference once its constraining effect on the charm mass has been taken into account.
Adding data on jet productuion also does not make much difference IF the value 
of $\alpha_s(M_Z^2)$ is kept fixed. However if  $\alpha_s(M_Z^2)$ is free then jet data have a 
dramatic effect in constraining its value, see Fig.~\ref{fig:d15039f65}, where
 the $\chi^2$ profiles vs $\alpha_s(M_Z^2)$ are shown for the NLO and NNLO fits 
to inclusive data alone and the same profile is shown for the NLO fit including
 jets. (Note that we cannot include jets in an NNLO fit since jet production cross sections in DIS have not been calculated to NNLO). 
A simultaneous fit of the PDFs and the value of $\alpha_s(M_Z^2)$ 
can be made once the jet data are included resulting in the value, 
$\alpha_s(M_Z^2) = 0.1183\pm 0.009(exp) \pm 0.0005(model/param) \pm 0.0012 (had) ^{+ 0.0037}_{-0.0030}(scale)$, where 'had' indicates extra uncertainties due to the hadronisation of the jets. 
The gluon PDF is strongly correlated to the value of $\alpha_s(M_Z^2)$ and 
thus, in a fit where  $\alpha_s(M_Z^2)$ is free, the gluon uncertainty increases.
 However, provided that 
jet data are included in the fit this increase is not dramatic, see Figs~\ref{fig:d15039f66af66b}.

\section{Conclusions}
The H1 and ZEUS data on inclusive $e^{\pm}p$ neutral and charged current cross sections have been
 combined into a data set with a total integrated luminosity of $\sim 1fb^{-1}$. This data set 
spans six orders of magnitude in both $x$ and $Q^2$. The combined cross sections were used as 
input to a pQCD analysis to extract the parton distribution functions HERAPDF2.0 at LO, NLO and NNLO. 
The effect of using various different heavy flavour schemes and different $Q^2$ cuts on the data 
was investigated. All heavy flavour schemes show some sensitivity to the minimim $Q^2$ cut, 
however the choice of this cut does not bias data at high scale significantly. For the standard fits the value of $\alpha_s(M_Z^2)$ is fixed, but a measurement of $\alpha_s(M_Z^2)$ can be made if jet data are included in the fit, resulting in the value
 $\alpha_s(M_Z^2) = 0.1184 \pm 0.0016$ at NLO, excluding scale uncertainties.
The data and the PDFs are available on https://www.desy.de/h1zeus/herapdf20.

\end{document}